# Damage Detection in Railroad Wheels from Impact-Induced Ultrasonic Measurements Using Self-Referenced Decay Analysis

Aashish Shaju, Nikhil Kumar, Steve Southward, and Mehdi Ahmadian

Center for Vehicle Systems and Safety (CVeSS), Virginia Tech

## Abstract

This study evaluates a non-contact screening method for stationary railroad wheelsets using air-coupled ultrasonic measurements excited by hammer impacts. Two decay features are considered. First, an effective damping ratio is obtained by normalizing the measured decay rate by the dominant response frequency. Second, strike-to-strike variations are measured to assess the quality of the energy-decay fit for the overall wheel. The method was evaluated using 22 wheelsets at two railroad facilities. The analysis indicates that a random forest trained on one dataset and tested on another performed poorly. This result motivated a simpler approach based on thresholds applied to the two decay features, which fixed the feature definitions and calibrated the thresholds on either dataset. This approach correctly classified every wheelset in the other dataset. A more stringent wheel-level validation repeated second-feature selection and threshold calibration while each test wheelset was withheld. This procedure correctly classifies 21 of 22 wheelsets. All defective wheelsets are identified, while one healthy wheelset yields a false positive. The results further show that the two-feature detector transferred between the tested datasets for stationary assessment of wheel damage. Further evaluation is required using worn but serviceable wheels, controlled support conditions, and additional independent datasets.

## 1. Introduction

Railway wheels develop tread and rim damage during service, and undetected damage can lead to wheel failure. An investigation of broken rims in North American freight service documented accidents involving shattered and vertically split rims and proposed a crack-growth sequence linking tread shelling and spalling to vertically split rims [1]. Wheel condition is assessed through workshop inspection, onboard monitoring, and wayside measurements [2]. Wheel impact load detectors measure the vertical force from passing wheels, and trends in these measurements have been used to identify wheels with a higher probability of failure [3]. Other wayside methods use stress, sound, or images to detect wheel flats, with performance affected by the sensing arrangement and operating conditions [4]. These measurements depend on the excitation produced during passage. Stationary impact testing allows the operator to select the excitation location and repeat a measurement when needed.

Acoustic measurements can provide information about wheel condition without placing a sensor on the wheel. In trackside recordings of light rail vehicles, wheel flats have been identified from the modulation and harmonics produced by repeated flat impacts [5]. Acoustic emission studies also show that the measured response depends on factors beyond damage alone. Tests on a scaled wheel-rail rig found changes in recorded energy with varying contact conditions [6]. In loaded rail specimens, emission was attributed to crack growth, contact between crack faces, and material-related sources also present in specimens without introduced cracks [7].

Air-coupled ultrasonic reception with laser excitation has been used to inspect rails and wheels, including flange and tread thermal-fatigue cracks and subsurface shattered-rim cracks [8]. Passive ultrasonic rail inspection using the rolling wheel as the source has also been demonstrated at speeds up to 80 mph, using a reconstructed transfer function and statistical outlier analysis [9]. These studies support the use of air-coupled sensors in railway inspection, although their excitation methods and inspected structures differ from the stationary wheel tests considered here.

Damage identification from structural response commonly uses statistical pattern recognition applied to response features [10]. One approach infers damage from changes in natural frequencies, treating damage as a local reduction in stiffness. Vestroni and Capecchi [11] examined the frequency measurements needed to locate and quantify damage in beams and beam systems, using both experimental and simulated data. Such approaches require a model or measurement of the undamaged structure. Damping is another damage-sensitive quantity. Impact tests on friction-stir-welded aluminum specimens showed changes in free decay and damping associated with defects confirmed by radiography and sectioned macrographs [12]. In wheel screening, an individual model or an earlier measurement of the same wheel is usually unavailable. A feature calculated from the current response would therefore be useful for wheels with different dimensions and service histories.

The performance of a response feature can change when it is applied to a different population. Population-based structural health monitoring addresses differences caused by environmental, operational, manufacturing, and structural properties through mappings between populations [13]. Studies of statistical alignment on numerical structures and the Z24 and KW51 bridges found that both the normalization method and class composition affect the outcome [14]. Feature selection also affects the reliability of a performance estimate. Selecting a feature using all available data and then evaluating it on a subset of that data can produce an optimistic estimate because selection itself involves fitting [15]. Repeating feature selection within each validation fold allows its performance to be evaluated using the withheld test samples.

Earlier work by the authors examined air-coupled ultrasonic responses from hammer-excited railway wheels and identified the ring-down decay rate as a useful discriminator in the populations tested [16]. A subsequent study applied a random forest to features from eleven wheelsets spanning nine health states [17]. Both studies evaluated performance within the population used for development. The present study examines transfer between two datasets acquired independently at different sites, comprising 22 wheelsets in total. A detector based on two response features is evaluated alongside a model using a larger feature set. The main contributions are:

- Definition of an effective damping ratio from a single impact and a variability feature from three repeated impacts at one position, including the validity checks applied to each feature.
- Evaluation of a two-feature threshold rule that combines position-level flags into a wheelset classification, with thresholds calibrated on one dataset and applied unchanged to the other in both directions.
- Assessment of second-feature selection through nested wheel-level validation and through selection restricted to a single dataset.

## 2. Measurement System and Test Campaigns

### 2.1 Requirements on the Acquisition System

A data acquisition system was built and used to capture ultrasonic ring-down and to assess signal quality during testing. Its main requirements were sufficient bandwidth and dynamic range, repeatable sampling timing, on-site processing, and traceable recordings.

The analysis extends to 80 kHz, and the microphone remains useful to about 120 kHz. The driver stage described in Section 2.2 provides the only analog anti-alias filtering, with a first-order corner near 55 kHz. For components that fold according to $|f_s - f|$, a sampling rate of 200 kSPS maps the 120 to 180 kHz interval into the 20 to 80 kHz analysis band. The front end attenuates this interval by about 8 to 11 dB, and the microphone still responds there. At 500 kSPS, the corresponding interval is 420 to 480 kHz, where attenuation is about 18 dB and no acoustic content of interest is expected. The higher sampling rate moves this alias interval away from the useful microphone range. The analog corner remains at 55 kHz, and its effect within the analysis band is compensated as described in Section 3.1.

Sampling is controlled in programmable logic to provide a fixed sample interval. Accurate timing and complete records are required because the decay rate is calculated from the change in response over time. Timing errors or missing samples would affect that estimate.

The available dynamic range must support both the signal-quality check and the decay fit. A strike must have a peak at least 25 dB above the noise floor, and the fit must span a 20 dB decrease in the energy decay curve. The measured system noise is approximately 20 μV RMS, corresponding to a dynamic range of 103.2 dB and about 17 effective bits. Battery power and a galvanically isolated Ethernet connection reduce ground-loop noise and allow the host computer to be placed away from the wheelset.

Data are streamed to a MATLAB host so that signal quality can be checked and a rejected strike repeated while the test crew is at the wheelset. Each recording includes the wheelset identifier, measurement position, and strike number. These identifiers preserve the grouping required for wheel-level and cross-dataset evaluations. The system uses documented components to support reproduction of the measurement setup.

## 2.2 Signal Chain

Figure 1 shows the acquisition chain and field configuration. A non-contact ultrasonic microphone is positioned approximately 330 to 355 mm from the wheel surface. Its output passes through a fully differential driver stage and a 24-bit successive-approximation converter operating at 500 kSPS per channel with simultaneous sampling. The driver's feedback network sets the low-pass corner near 55 kHz. The converter has no decimation filter, so the driver stage provides the only anti-alias filtering. Programmable logic on a system-on-chip development board controls acquisition, and strike records are buffered and transferred to the host over Ethernet. Both microphones are oriented axially toward the wheel plate. The analysis in this paper uses Channel 1.

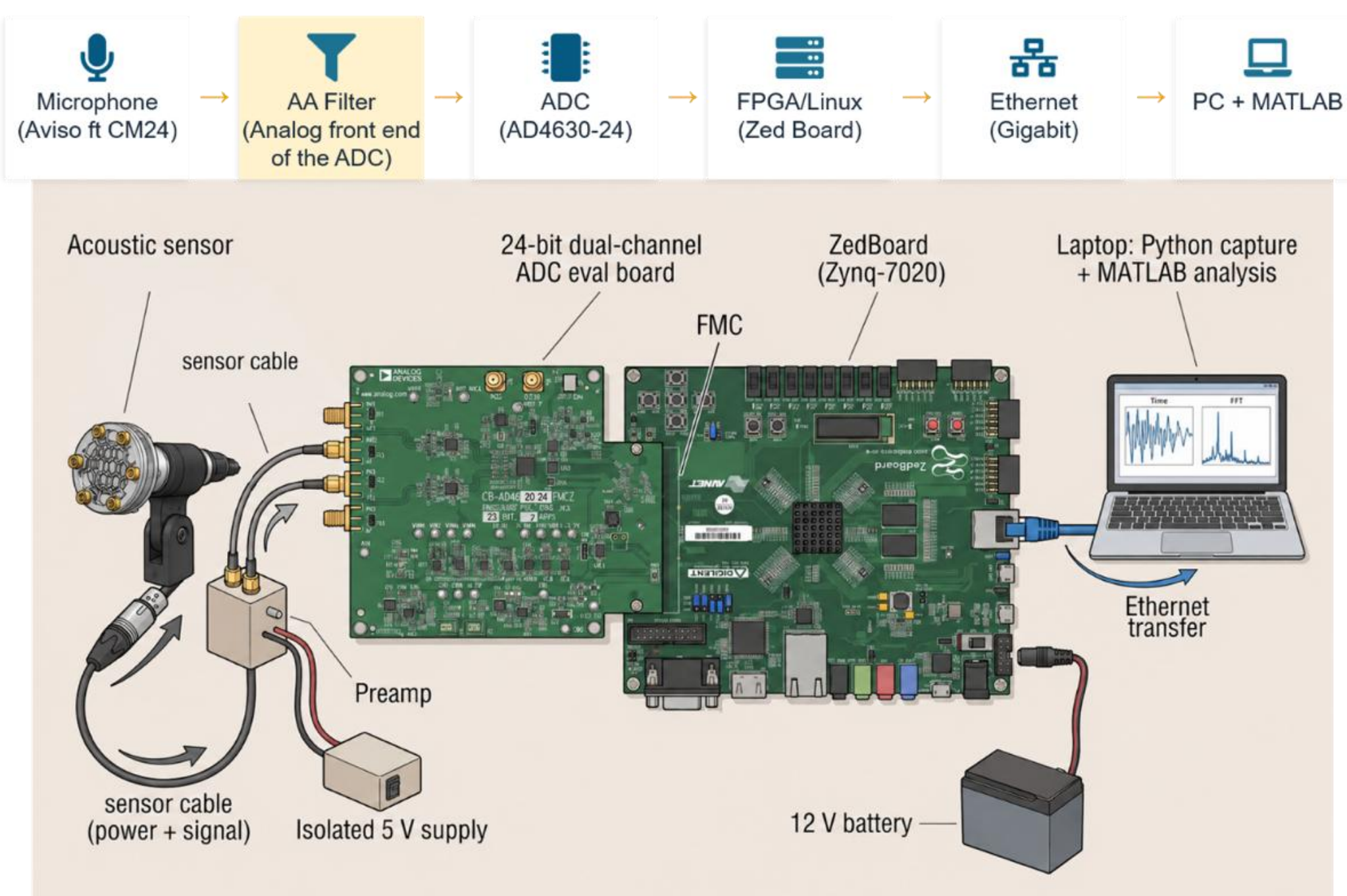

Figure 1. Air-coupled ultrasonic measurement system used for the wheelset impact tests, showing the microphone, analog front end, 24-bit analog-to-digital converter, FPGA-based acquisition board, Ethernet connection and host computer.

## 2.3 Wheelsets, Excitation Protocol and Datasets

Eleven wheelsets were measured at the MxV Rail facility in Pueblo, Colorado, and eleven more at the Norfolk Southern Railroad facility in Roanoke, Virginia. The two independent datasets contain 22 wheelsets in various conditions, which we identified using nine labels. The MxV set included two healthy and nine defective wheelsets, while the Norfolk Southern set had three healthy and eight defective wheelsets. All five healthy wheelsets were new. Figure 2 shows the wheelsets and examples of labels for each condition.

Each wheelset was struck at six positions [16], with three strikes per position, yielding 396 strikes and 132 wheel-position records per channel. The positions were on the tread and flange at the three, nine, and twelve o'clock orientations. The response was measured without contact, with no transducer mounted on the wheel.

All MxV wheelsets were measured while standing on the rail. At Norfolk Southern, the healthy wheelsets stood on the rail, and the defective wheelsets rested on the ground. Therefore, support condition coincides with class in the Norfolk Southern dataset. Section 6.4 discusses the resulting limitation.

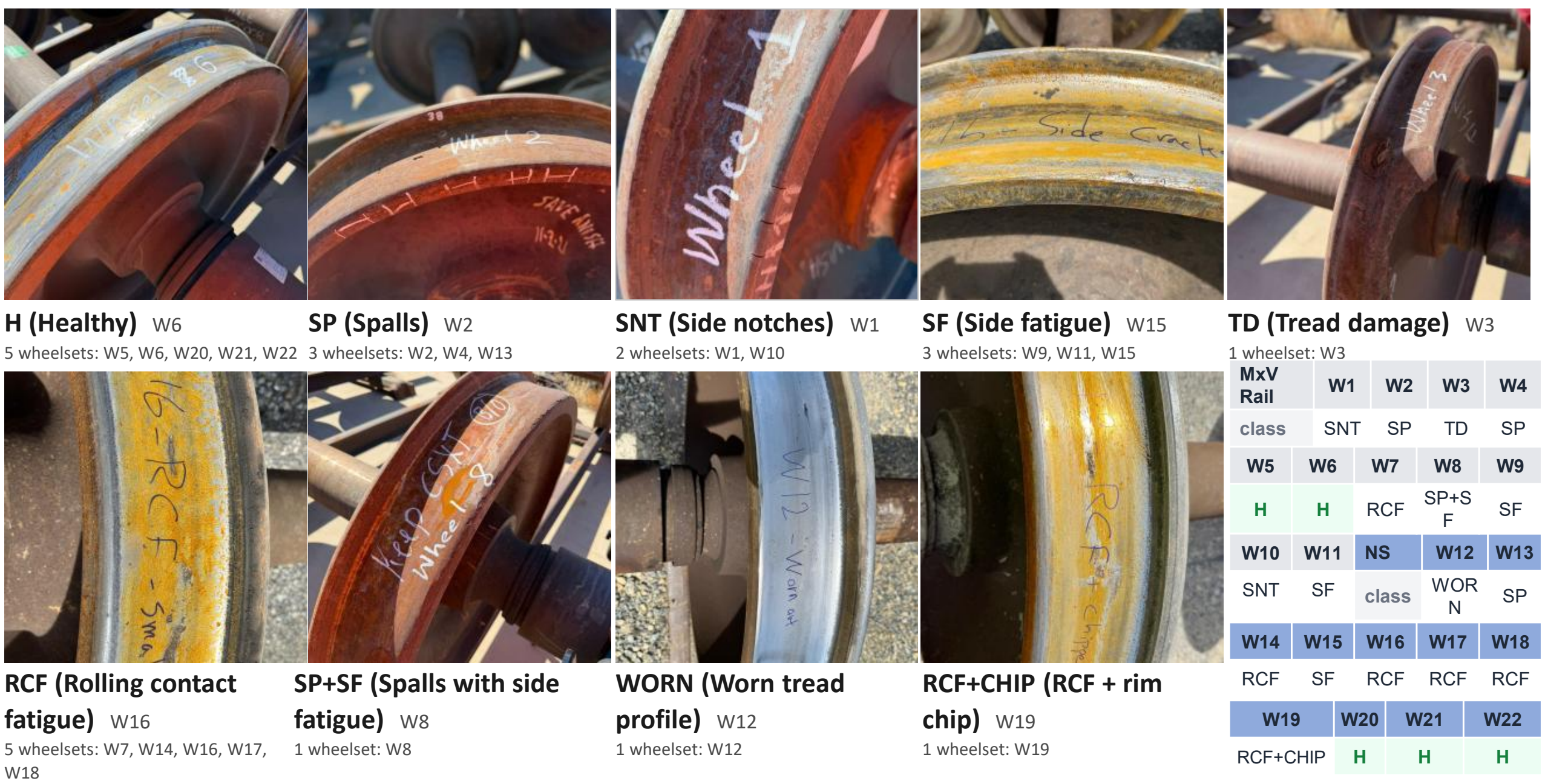

Figure 2. Wheelsets included in the MxV and Norfolk Southern datasets, grouped by condition label. One representative wheelset is shown for each condition, and the matrix on the right lists the assigned condition of wheelsets W1–W22.

# 3. Signal Processing and Feature Definitions

## 3.1 Front-End Compensation and Band Selection

Every strike is corrected for the measured analog front-end response before features are calculated. The driver stage has a fixed, repeatable response with a low-pass corner near 55 kHz, within the analysis band. Its response was measured using a swept sine. The measured magnitude is converted to a minimum-phase representation using a cepstral method [18], and the inverse is formed via Wiener-type deconvolution.

$$G(f) = H^*(f) \,/\, (\,|H(f)|^2 + \lambda\,) \qquad (1)$$

In Eq. (1), H(f) represents the front-end response, and λ is a frequency-dependent regularization term referenced to the noise floor. This regularization limits inverse gain when the front-end response is small, thereby reducing noise amplification at high frequencies. The correction recovers amplitude to within approximately one percent up to 120 kHz. All subsequent calculations use the compensated pressure signal p(t).

Figure 3 shows the front-end response before and after compensation. The analysis uses the 20 to 80 kHz range, where the microphone captures measurable wheel ring-down. This range is divided into four sub-bands: 20 to 35 kHz, 35 to 50 kHz, 50 to 65 kHz, and 65 to 80 kHz. Two processing branches are used. For the damping feature, the compensated record is restricted to the selected sub-band, and the energy decay curve, decay slope, and dominant frequency are then calculated. For the variability feature, the energy decay curve and EDR_R2 are calculated from the compensated 20 to 80 kHz record.

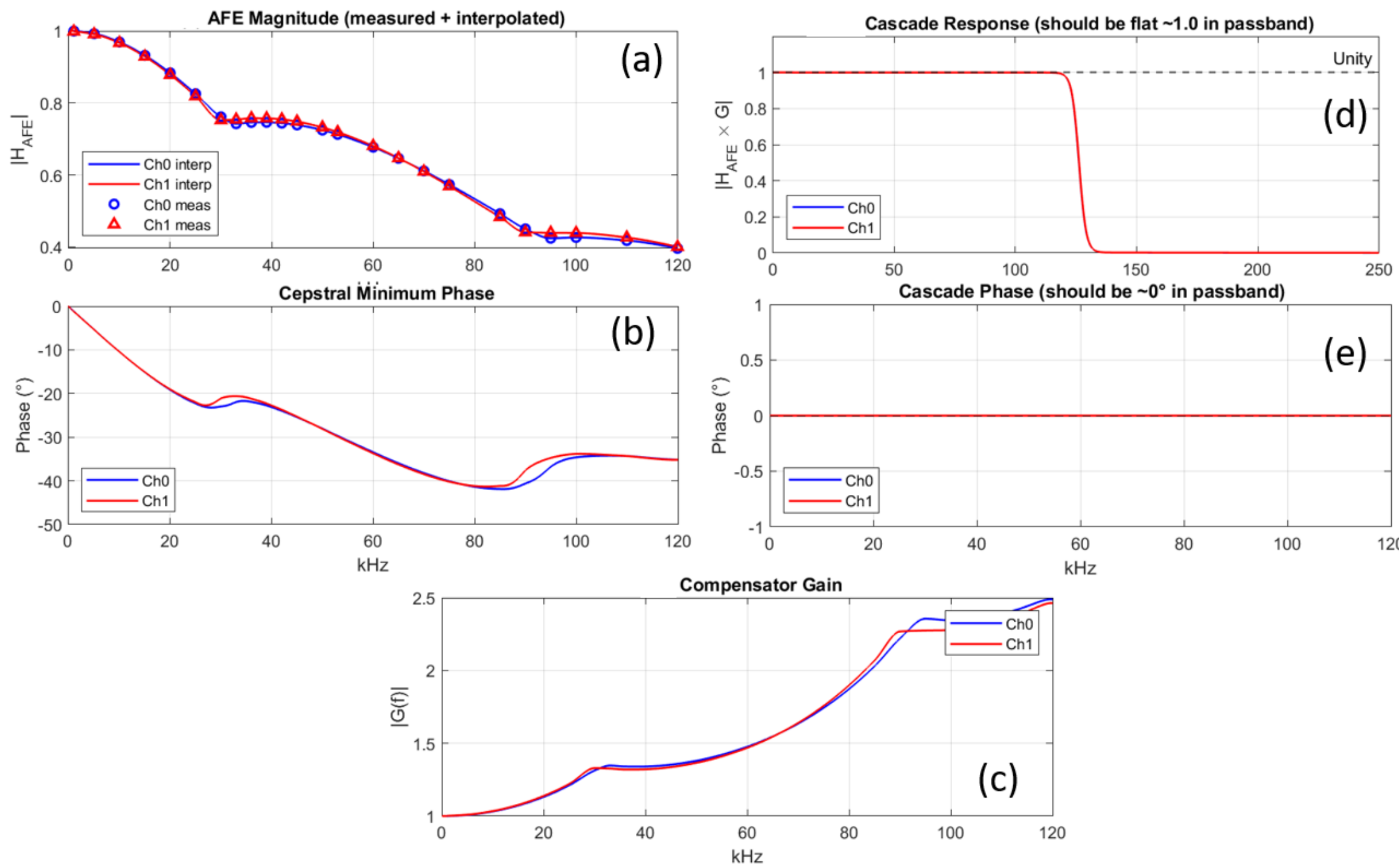

Figure 3. Measured analog-front-end response and the resulting minimum-phase compensation: (a) measured and interpolated magnitude response, (b) reconstructed minimum-phase response, (c) compensator gain, (d) compensated cascade magnitude and (e) compensated cascade phase.

## 3.2 Energy Decay Curve and Noise-Floor Truncation

For each processing branch, the decay is computed from the Schroeder integral of the squared, compensated pressure signal [19]. In a response with multiple modes, beating causes the amplitude envelope to fluctuate, making an envelope fit sensitive to the chosen time window. Backward integration yields a monotonically decreasing energy curve and reduces this sensitivity. Figure 4 illustrates the calculation for one strike. For a record beginning at the response peak,

$$E(t) = \int p^2(\tau)\, d\tau\,, \quad t \le \tau \le t_c \qquad (2)$$

$$L(t) = 10 \log_{10} [\, E(t) / E(0) \,] \qquad (3)$$

The truncation time $t_c$ in Eq. (2) is determined using the Lundeby procedure [20]. The procedure begins with a noise-floor estimate from the final 0.30 s of the record and computes short-time energy averages in 10 ms blocks. It estimates the crossover between the decaying response and the stationary noise floor using a 5 dB margin above the floor. The first 0.05 s after the crossover is excluded when updating the noise estimate. The calculation is repeated until convergence or a maximum of eight iterations. Truncation limits the noise energy included in the integral, which would otherwise flatten the decay curve and bias the fitted rate. The effect of stationary noise on decay estimation has also been examined by Karjalainen et al. [21]. The Lundeby truncation time and the fitted decay interval were recalculated separately for each frequency branch.

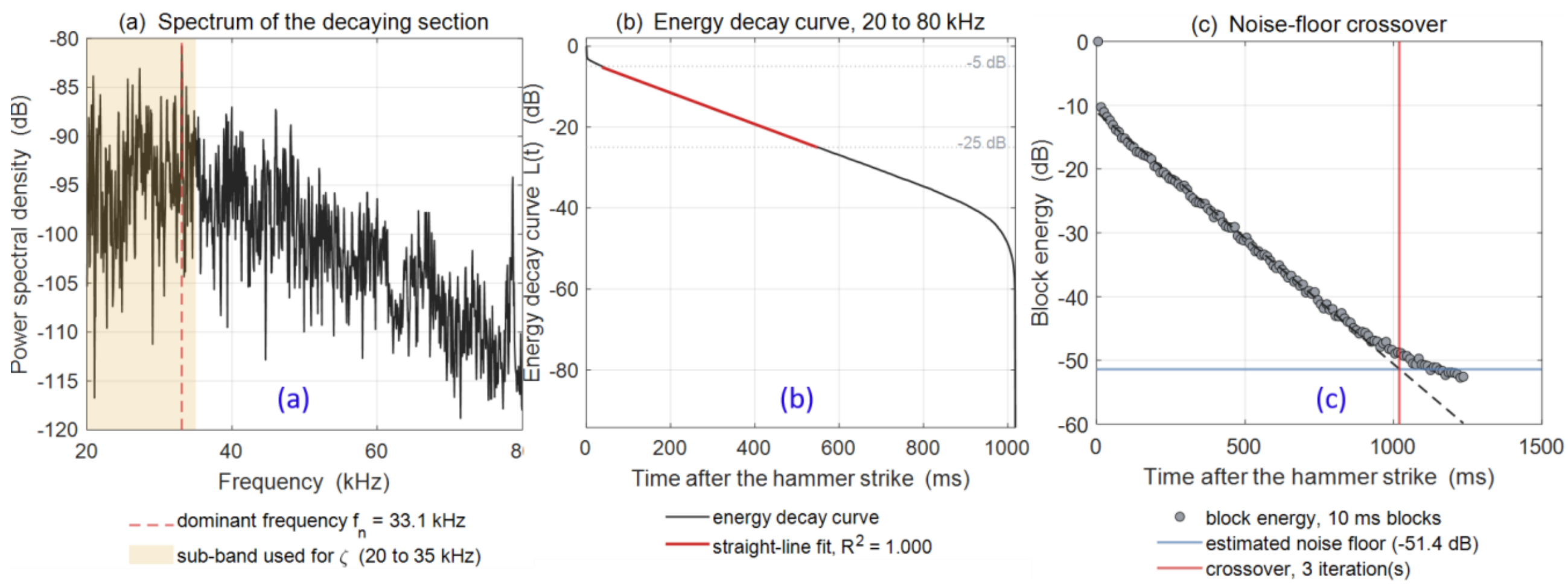

Figure 4. Processing of one hammer strike: (a) spectrum of the decaying response and dominant frequency in the 20–35 kHz band, (b) broadband energy decay curve and straight-line fit and (c) Lundeby noise-floor crossover used to determine the end of the decay interval.

## 3.3 Frequency Normalization and the Effective Damping Ratio

A straight line is fitted to the 20 to 35 kHz band-limited decay curve $L(t)$ between 5 dB and 25 dB below its initial value. This window excludes the initial transient and limits the contribution from the region near the noise floor. The fitted slope m is converted to an amplitude decay constant,

$$\alpha = |m| \,/\, 8.6859 \quad [s^{-1}], \qquad (4)$$

which is normalized by the dominant response frequency within the same band to obtain the effective damping ratio,

$$\zeta = \alpha \,/\, (2\pi f_n) \qquad (5)$$

The measured decay rate depends partly on how fast the wheel is vibrating. For the same level of damping, a higher-frequency response completes more vibration cycles per second and therefore decays faster when the rate is expressed per second. Equation (5) reduces this frequency effect by dividing the decay constant (α) by the angular frequency, $2\pi f_n$. Here, $f_n$ is the frequency of the largest peak in the measured ring-down spectrum within the selected band. It is obtained directly from the measured signal and is not a modal frequency calculated from a wheel model. Therefore, ζ is described as an effective damping ratio rather than a modal damping ratio. Both α and fn are calculated from the same hammer strike. No separate baseline test is used in this calculation. Section 5.4 examines whether a threshold calibrated using one dataset can be applied to the other dataset.

A damping estimate is retained only when the signal is sufficiently strong, and the decay is adequately represented by an exponential curve. First, the peak of the complete signal must be at least 25 dB above the broadband noise floor. Second, the dominant peak within the selected frequency band must be at least 22 dB above the noise floor in that band. A straight line is then fitted to the energy decay curve. The fit must have a coefficient of determination, $R^2$, of at least 0.95. This requirement assesses whether a single exponential reasonably describes the measured decay over the selected interval. If either signal-strength requirement is not met, or if the fitted line has an $R^2$ below 0.95, the damping estimate is considered unreliable and is recorded as missing.

The detector uses ζ from the 20–35 kHz sub-band. The damping ratios calculated in the other three sub-bands did not separate the condition classes and were therefore excluded from the decision rule.

### 3.4 Strike-to-Strike Energy Decay-Fit Variability

At each wheel position, the test includes three repeated hammer strikes. For each strike, EDR_R2 quantifies how closely the broadband 20 to 80 kHz energy decay curve follows a straight line on a decibel scale. A value close to one indicates that the decay is well represented by a single exponential. The variability feature measures the spread of the retained EDR_R2 values across repeated strikes and is calculated as the interquartile range of EDR_R2 at a single position.

$$EDR_R2_iqr = IQR\{\ EDR_R2\ \} \qquad (6)$$

A small variability value indicates that the decay curves have similar fit quality across strikes. A large value indicates that at least one strike yields noticeably different fit quality. This may suggest that the measured response is not consistent between impacts, although the feature does not identify the physical cause of that difference. Figure 5 compares a healthy wheelset with a defective wheelset whose effective damping ratio remains within the healthy range. The example shows how the variability feature can identify a difference that the damping feature does not capture.

The two features are computed from the same recordings but capture different aspects of the response. The damping feature uses the decay slope and the dominant frequency in the 20 to 35 kHz band. It quantifies the rate of energy loss relative to the response frequency. The variability feature uses the fit quality of the broadband 20 to 80 kHz energy decay curve and measures how much that fit quality varies across repeated strikes. Both features are dimensionless.

The signal-quality and fit-quality checks serve different purposes. The signal-quality screen verifies that the recorded response is sufficiently above the measurement noise. Every strike used by either feature must have a broadband peak at least 25 dB above the broadband noise floor. A damping estimate must also have an in-band peak at least 22 dB above the noise floor in the 20 to 35 kHz band. The fit-quality check then assesses whether the band-limited energy decay curve is adequately represented by a single exponential. As described above, the corresponding straight-line fit must have an $R^2$ of at least 0.95 for the damping estimate to be retained.

The variability calculation treats the fit-quality check differently. If a strike passes the signal-quality screen but its decay fit has an $R^2$ below 0.95, the damping estimate is rejected, while the broadband EDR_R2 value is retained for the variability calculation. This is intentional because the variability feature measures differences in fit quality across repeated strikes. Removing every strike with a poor fit would eliminate part of the variation the feature is intended to detect.

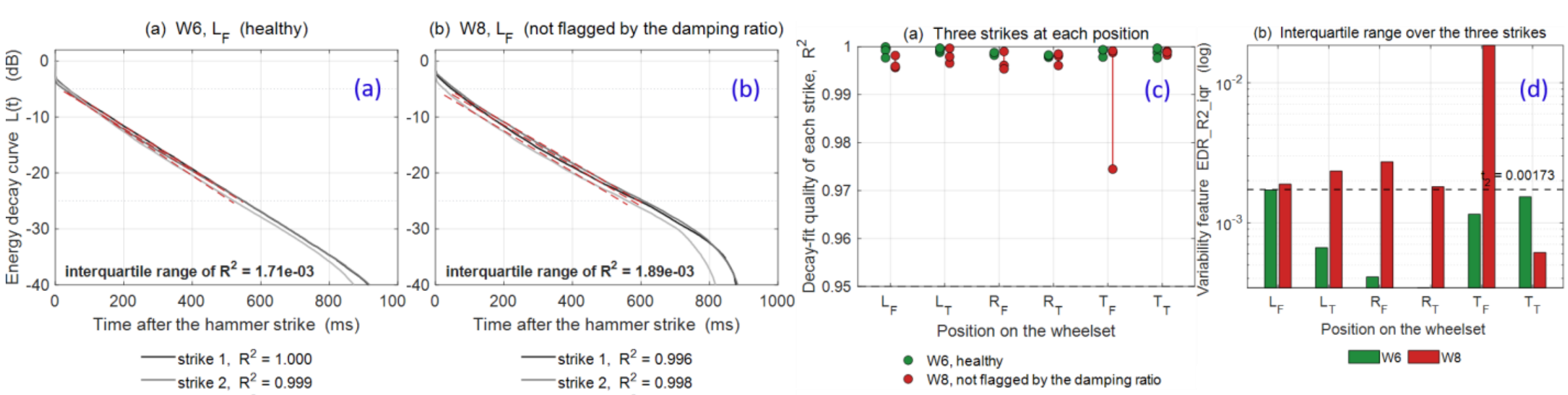

Figure 5. Calculation of strike-to-strike energy decay-fit variability. Panels (a) and (b) show the broadband energy decay curves and fitted lines for three strikes on healthy wheelset W6 and defective wheelset W8. Panel (c) compares the decay-fit quality across the six measurement positions, and panel (d) shows the resulting position-level variability. W8 lies within the healthy damping range but shows greater variation in decay-fit quality.

### 3.5 Aggregation and Decision Rule

The damping feature at each position is the median of the valid $\zeta$ estimates from the three strikes. The variability feature is the interquartile range in (6), calculated from the retained EDR_R2 values. It is computed when two or three values remain and is marked missing when fewer than two remain. These summaries yield one damping value and one variability value per position.

A position is flagged when its effective damping ratio exceeds $t_1$ or its variability feature exceeds $t_2$.

$$flag(position) = [\, \zeta > t_1 \,] \;\vee\; [\, EDR_R2_iqr > t_2 \,] \tag{7}$$

A wheelset is classified as defective when at least three of its six positions are flagged, including in the case of a three-to-three tie. When fewer than six positions are available, a strict majority of the available positions must be flagged. If one feature is missing at a position, the decision uses the available feature. If both are missing, the position is counted as flagged. This treatment allows insufficient measurements to contribute to referral for inspection.

Figure 6 presents the complete decision rule. The OR combination flags a position if either feature detects a response pattern, since the features capture different response patterns. Both thresholds are calibrated using only the training data, as described in Section 4. We evaluate feature selection separately from threshold calibration.

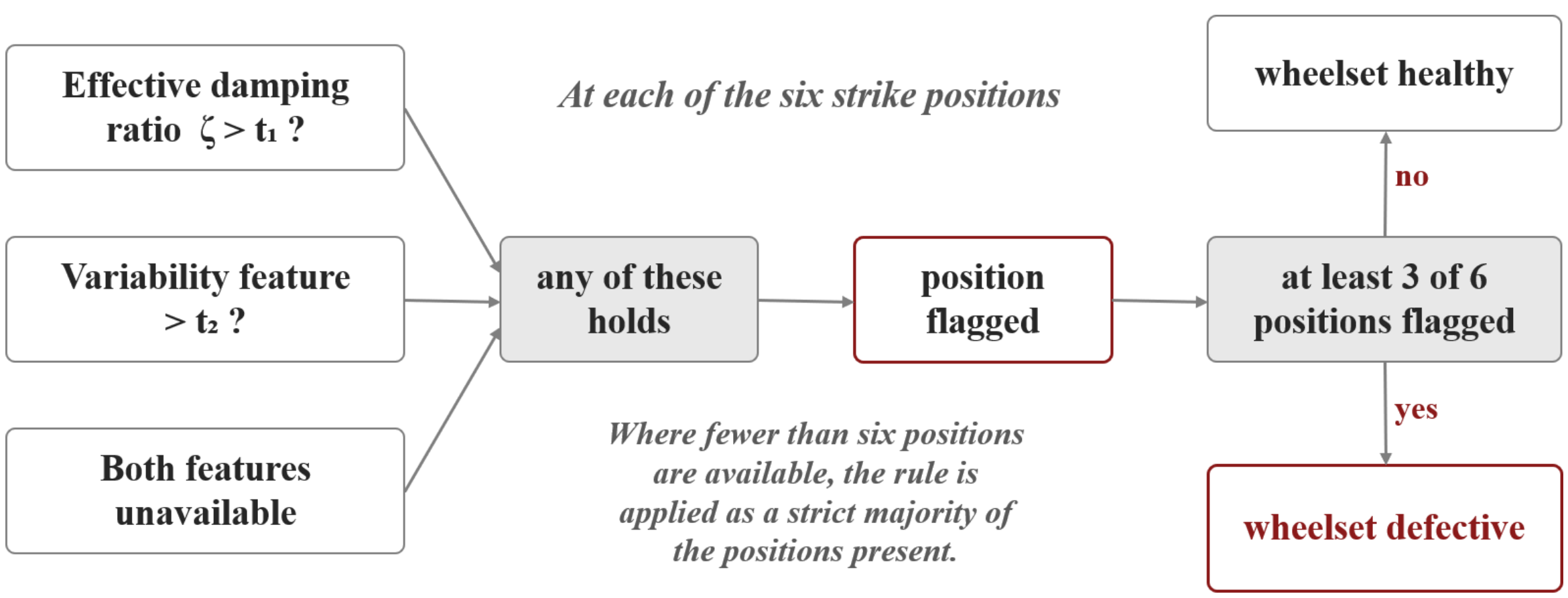

Figure 6. Decision rule from position-level feature values to wheelset classification.

## 4. Evaluation Design

The detector is developed in two steps. First, the second feature is selected from a group of candidate features. Second, a threshold is calculated for each of the two selected features using the training data. The same method is used to calculate the thresholds in every evaluation. However, each validation scheme places different restrictions on which wheelsets can be used for feature selection and threshold calibration. The remaining wheelsets are kept separate for testing. This distinction allows the study to evaluate both whether the thresholds transfer between datasets and whether the same second feature would be selected when the test data are not available during development.

## 4.1 Threshold Calibration

A separate threshold is calculated for each feature using the training wheelsets. For each feature, the available values from all training positions are sorted in increasing order. Trial thresholds are then placed midway between each pair of consecutive values. Missing values are not used to define these trial thresholds.

Each trial threshold is evaluated by applying it to the training positions. A position is flagged if its feature value exceeds the threshold. During this single-feature calibration step, a missing value does not trigger a flag. This treatment applies only during the evaluation of a single feature's threshold. Missing values in the final two-feature detector are handled according to the rules in Section 3.5.

The performance of each trial threshold is measured by position-level balanced accuracy,

$$BA_{pos} = ½\,[\,TNR_{healthy} + TPR_{defective}\,], \qquad (8)$$

where $TNR_{healthy}$ is the proportion of healthy training positions that remain unflagged, and $TPR_{defective}$ is the proportion of defective training positions that are correctly flagged. These two values are averaged to give equal weight to healthy and defective wheels, even when they have different numbers of positions. We select the threshold with the highest balanced accuracy. If several thresholds yield the same balanced accuracy, we use the lowest of those thresholds.

This procedure is carried out independently for the damping and variability features. Calibration is performed at the position level so that every usable training position contributes to selecting the thresholds. After the two thresholds are selected, they are applied together using the decision rule in Section 3.5. The position flags are then combined to classify the complete wheelset. Therefore, a single flagged position does not, by itself, cause a wheelset to be classified as defective.

## 4.2 Validation Schemes

Figure 7 summarizes the three main validation schemes. Each scheme addresses a different question by varying which wheelsets are used for training and which are withheld for testing. Accordingly, the results are reported separately.

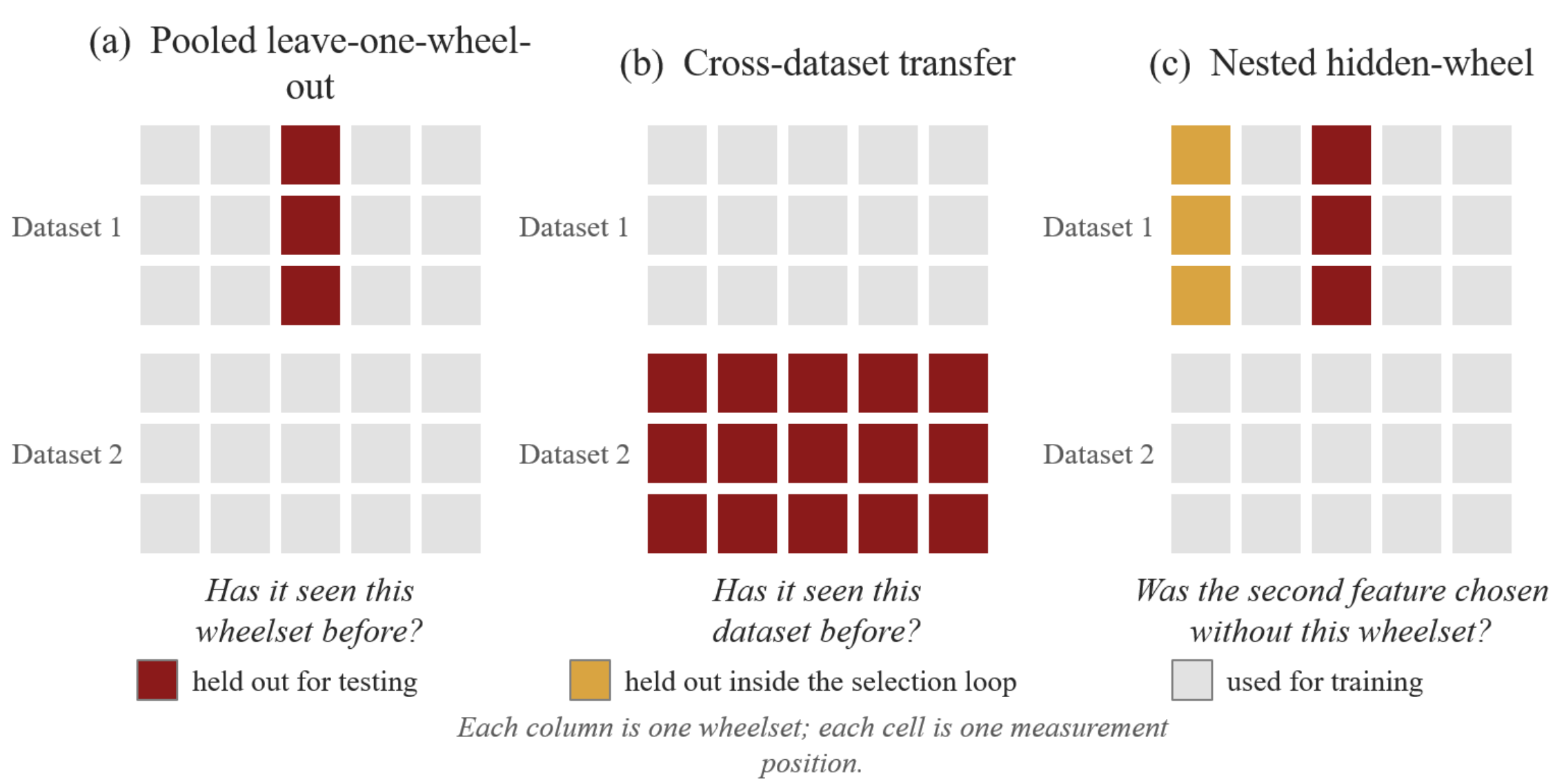

Figure 7. The three validation schemes and what each of them holds out.

In pooled leave-one-wheel-out validation, the two datasets are combined. In each round, one complete wheelset, including all its measurement positions, is withheld for testing. The thresholds are calculated from the remaining 21 wheelsets and then applied to the withheld wheelset. The test wheelset does not influence threshold calibration. However, wheelsets from both sites remain in the training data.

Cross-dataset transfer evaluates whether thresholds obtained at one site can be applied at another. The thresholds are calibrated using all wheelsets from one dataset and then applied without modification to all wheelsets in the other dataset. The process is repeated in the opposite direction. The two features remain fixed because they were selected during development using both datasets. Therefore, this evaluation tests the transfer of the thresholds, not the independent selection of the features.

Nested hidden-wheel validation also withholds one complete wheelset in each round. In this case, the withheld wheelset is excluded from both selecting the second feature and calibrating the thresholds. The remaining 21 wheelsets are used to select the second feature and to calculate both thresholds. The resulting detector is then applied to the withheld wheelset. This process is repeated until every wheelset has been tested. The damping feature, the list of candidate second features, and the wheelset decision rule remain fixed. Repeating feature selection without the test wheelset reduces the selection bias that can arise when the same data influence both detector development and evaluation [22].

An additional evaluation performs both second-feature selection and threshold calibration using only one dataset, then tests on the other. The process is then repeated in the opposite direction. The damping feature is also evaluated by itself within the training dataset. If damping already separates all the training wheelsets, the selection procedure has little information to determine which second feature will transfer best. Section 4.3 describes this selection procedure, and Section 5.6 presents the results. Because the final two-feature detector was developed using information from both current datasets, a fully independent evaluation will require a third dataset.

### 4.3 Selection of the Second Feature

The effective damping ratio is always used as the first feature and is not reselected during nested validation. Its position-level value remains the median effective damping ratio in the 20 to 35 kHz band. Only the second feature is selected. Each candidate is tested using an inner leave-one-wheel-out procedure on the available training wheelsets. In each inner round, one training wheelset is withheld. The thresholds for the damping feature and the candidate feature are calibrated using positions from the remaining wheelsets, following Section 4.1. The two features are then combined using the OR rule, and the withheld wheelset is classified using the position-count rule in Section 3.5. This process is repeated until every training wheelset has been withheld once. The score for a candidate is the balanced accuracy of its wheelset classifications across all inner rounds. Therefore, the thresholds are calibrated using position-level performance, while the second feature is selected using wheel-level performance.

Candidates are ranked first by balanced accuracy and then by healthy-class recall. If candidates remain tied, their order in the original feature table serves as the final tie-breaker. The number of candidates tied for the highest balanced accuracy is also recorded. A large number of tied candidates indicates that the available training wheelsets do not clearly distinguish among them. The damping feature is evaluated in isolation using the same inner procedure. This diagnostic determines whether damping already separates the training wheelsets before a second feature is added. An inner balanced accuracy of 0.98 or higher is treated as indicating that damping already provides nearly complete separation of the training set.

### 4.4 Metrics

Performance is reported using healthy-class recall, defective-class recall, and balanced accuracy. Healthy-class recall is the proportion of healthy wheelsets correctly classified as healthy. Defective-class recall is the proportion of defective wheelsets correctly classified as defective. Balanced accuracy is the average of these two recall values, so the healthy and defective classes receive equal weight. This distinction is important because the dataset contains five healthy and seventeen defective wheelsets. A detector that labels every wheelset as defective would correctly classify 17 of 22 wheelsets and achieve a plain accuracy of 0.77. However, it would have zero healthy-class recall and a balanced accuracy of only 0.50. Balanced accuracy therefore reveals poor performance on the smaller healthy class that would be hidden by plain accuracy.

Table 1 reports the metrics at the level used in each evaluation, either the position or wheelset level. Wheelset-level results are computed by combining the position flags according to the rule in Section 3.5. The table also reports the number of correctly classified wheelsets and identifies each misclassified wheelset.

# 5. Results

## 5.1 Pooled Evaluation of the 109-Feature Model

The 109-feature random forest [23] used 500 trees and was evaluated using 22 leave-one-wheel-out folds. Scaling and class weighting were fitted using only the training wheelsets. Performance was scored at the position level across the six positions of each withheld wheelset. Healthy-class recall was 0.73, defective-class recall was 0.79, and balanced accuracy was 0.76 (Figure 8(a)). Both datasets were represented in every training fold.

## 5.2 Cross-Dataset Evaluation

The model described above was trained on eleven MxV wheelsets and then tested on eleven Norfolk Southern wheelsets. Norfolk Southern data were not used during training. At the position level, the average healthy-class recall across twenty independently seeded models was only 0.06. Defective-class recall was 0.90, yielding a balanced accuracy of 0.48. The standard deviation of each metric was 0.01 or less, indicating that the results changed very little between model runs. Only one of the eighteen healthy positions was classified correctly. The model therefore flagged nearly all healthy positions as well as most defective positions.

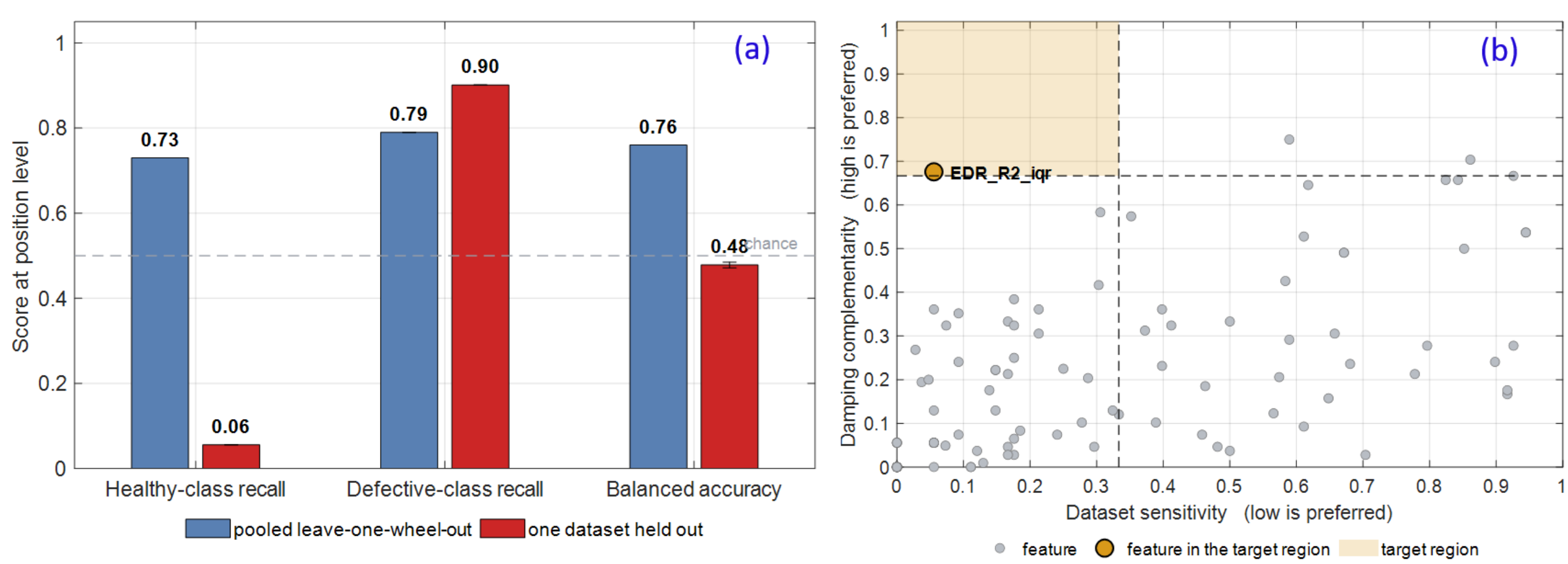

Figure 8. Figure 8. The 109-feature model. (a) Balanced accuracy, healthy-class recall, and defective-class recall under pooled leave-one-wheel-out validation and with one complete dataset held out. (b) Two-axis feature audit of all 109 features, with the target region shaded.

After the position-level results were combined into wheelset classifications, eight of the eleven Norfolk Southern wheelsets were classified correctly. The three errors were the healthy wheelsets W20, W21 and W22. At the position level, defective-class recall was 1.00 for the wear-condemned and rolling-contact-fatigue-plus-chip conditions, 0.92 for rolling contact fatigue, 0.83 for side fatigue and 0.67 for spalls. Although some individual defective positions were missed, all defective wheelsets were correctly classified after the position flags were combined. Therefore, the wheelset-level errors were limited to the three healthy wheelsets.

The test was repeated after removing 24 predictors from the decay-time, decay-rate, and dominant-frequency families. This reduced the model from 109 to 85 predictors while retaining the damping-ratio predictors. The results did not change to two decimal places: healthy-class recall remained 0.06, defective-class recall remained 0.90, and balanced accuracy remained 0.48. The same eight wheelsets were classified correctly. Therefore, the poor cross-dataset performance could not be corrected simply by removing these frequency-related predictors. This result motivated the development of a simpler detector using a small number of clearly defined response features.

The two datasets also contained different response patterns. In the Norfolk Southern dataset, every defective wheelset had a higher effective damping ratio than every healthy wheelset. The damping feature therefore separated the two classes in that dataset. In the MxV dataset, however, defective wheelsets W7, W8 and W9 had damping values within the healthy range. Changing the damping threshold alone could not distinguish these wheelsets without also affecting the healthy classifications. A second feature was therefore needed to identify this response pattern.

## 5.3 The Two-Axis Feature Audit

The 109 candidate features were compared using two criteria. The first criterion was that a feature should change as little as possible between the two measurement sites. The second criterion was that it should distinguish defective MxV wheelsets that the damping feature did not identify. We measured separation between the two groups using the score $2|\mathrm{AUC} - 0.5|$. The AUC was calculated from the ranked feature values using midranks to account for tied values. The resulting score ranges from zero to one. Midranks are defined as the average rank within a group of tied values. A score near zero indicates that the two groups have similar rankings and are not clearly separated. A score near one indicates strongly separated rankings. These scores are computed directly from the feature values and do not require a fitted threshold.

The first score, called dataset sensitivity, compares healthy positions from Norfolk Southern with healthy positions from MxV. A low dataset-sensitivity score is preferred because a useful transferable feature should yield similar values for healthy wheelsets at both sites. The second score, called damping complementarity, compares positions from defective wheelsets W7, W8 and W9 with those from healthy MxV wheelsets. A high complementarity score is preferred because W7, W8 and W9 fall within the healthy range of the damping feature. The second feature must therefore distinguish these wheelsets using a different part of the measured response. A score was calculated only when each comparison group contained at least five valid positions.

A candidate entered the target region when its dataset-sensitivity score was no greater than 1/3 and its damping-complementarity score was at least 2/3. It also had to produce a consistent ordering in at least five of the six wheel-level median comparisons between defective wheelsets W7, W8, and W9 and healthy wheelsets W5 and W6. This additional requirement checks that the position-level separation is supported by several wheelsets rather than being driven by a small number of individual positions.

The variability feature was the only one of the 109 candidates that met all the target-region requirements, as shown in Figure 8(b). Its dataset-sensitivity score was 0.06, indicating minimal separation between the healthy positions from the two sites. Its damping-complementarity score was 0.68, corresponding to an AUC of approximately 0.84, indicating useful separation of W7, W8 and W9 from the healthy MxV wheelsets.

## 5.4 Cross-Dataset Transfer of the Two-Feature Detector

This evaluation tested whether thresholds calibrated on one dataset could be applied directly to the other dataset. The two thresholds were calculated from one complete dataset and applied without modification to the wheelsets in the other dataset, as shown in Figure 9(a) and (b).

Calibration with the MxV dataset yielded a damping threshold of $t1 = 3.198 \times 10^{-5}$ and a variability threshold of $t2 = 1.727 \times 10^{-3}$. Calibration with the Norfolk Southern dataset yielded $t1 = 3.159 \times 10^{-5}$ and $t2 = 1.290 \times 10^{-3}$. The detector correctly classified all eleven test wheelsets in both transfer directions. The damping thresholds obtained independently from the two datasets differed by approximately one percent.

The two features identified distinct response patterns. Defective wheelsets W7, W8, and W9 had damping values within the healthy range and were therefore not flagged by the damping feature. However, their decay-fit quality varied more between repeated strikes, causing them to be flagged by the variability feature. Wheelset W19 showed the opposite pattern. Its decay fits were consistent across repeated strikes, so it was not flagged by variability, but its elevated damping ratio caused it to be flagged by the damping feature. As a result, every defective wheelset was identified by at least one of the two features.

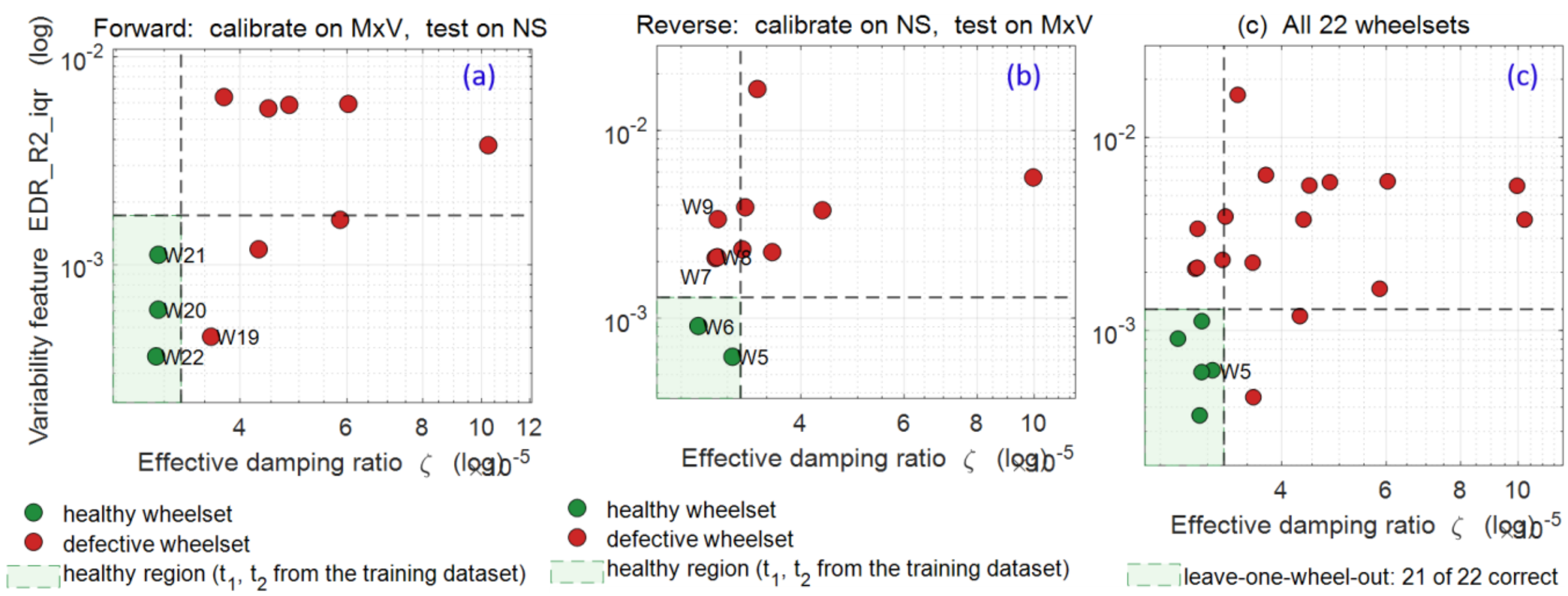

Figure 9. Wheelsets in the effective-damping-ratio and variability-feature plane. (a) Thresholds calibrated on MxV and applied to Norfolk Southern. (b) Thresholds calibrated on Norfolk Southern and applied to MxV. (c) All 22 wheelsets, with thresholds fitted to the complete dataset for reference. The reported pooled leave-one-wheel-out classifications use thresholds recalibrated in each fold and the position count in Section 3.5.

The contribution of the variability feature differed across the two transfer directions. When the detector was calibrated with MxV and tested on Norfolk Southern, the damping feature alone correctly classified all eleven Norfolk Southern wheelsets. The variability feature did not change these classifications. When the detector was calibrated with Norfolk Southern and tested on MxV, damping alone correctly classified only eight of the eleven MxV wheelsets. The variability feature identified W7, W8 and W9, increasing the result from eight of eleven to eleven of eleven. The second feature was therefore necessary for successful transfer in the Norfolk Southern-to-MxV direction.

## 5.5 Nested Hidden-Wheel Validation

Nested leave-one-wheel-out validation tests whether the same feature would be selected when the test wheelset is hidden. In each fold, one wheelset is withheld from both second-feature selection and threshold calibration. The remaining 21 wheelsets are used to select the second feature and calculate the thresholds, after which the detector is applied to the withheld wheelset. The damping feature, candidate family, and wheelset decision rule remain fixed throughout the evaluation.

Figure 10 compares the resulting classifications. The damping feature alone correctly classified 18 of the 22 wheelsets. It failed to identify the defective wheelsets W7, W8, and W9 and incorrectly flagged the healthy wheelset W5 as defective. When the second feature was selected separately within each fold, 21 of the 22 wheelsets were classified correctly, yielding a balanced accuracy of 0.90. All seventeen defective wheelsets were identified, and W5 was the only remaining error. The evaluation was also repeated with the variability feature fixed in advance, while both thresholds were recalibrated in each fold. This produced the same classifications, with a healthy-class recall of 0.80 and a defective-class recall of 1.00.

The variability feature was selected in 21 of the 22 folds. Only the fold in which W5 was withheld selected a different feature, DecayRate_iqr. In that fold, W6 was the only healthy MxV wheelset available during training. However, selecting a different second feature was not the cause of the W5 error. W5 was already classified as defective by the damping feature alone under the same fold calibration. Because the detector uses an OR rule, the second feature can add position flags but cannot remove flags produced by the damping feature. W5 therefore remained misclassified with both the selected second feature and the fixed variability feature.

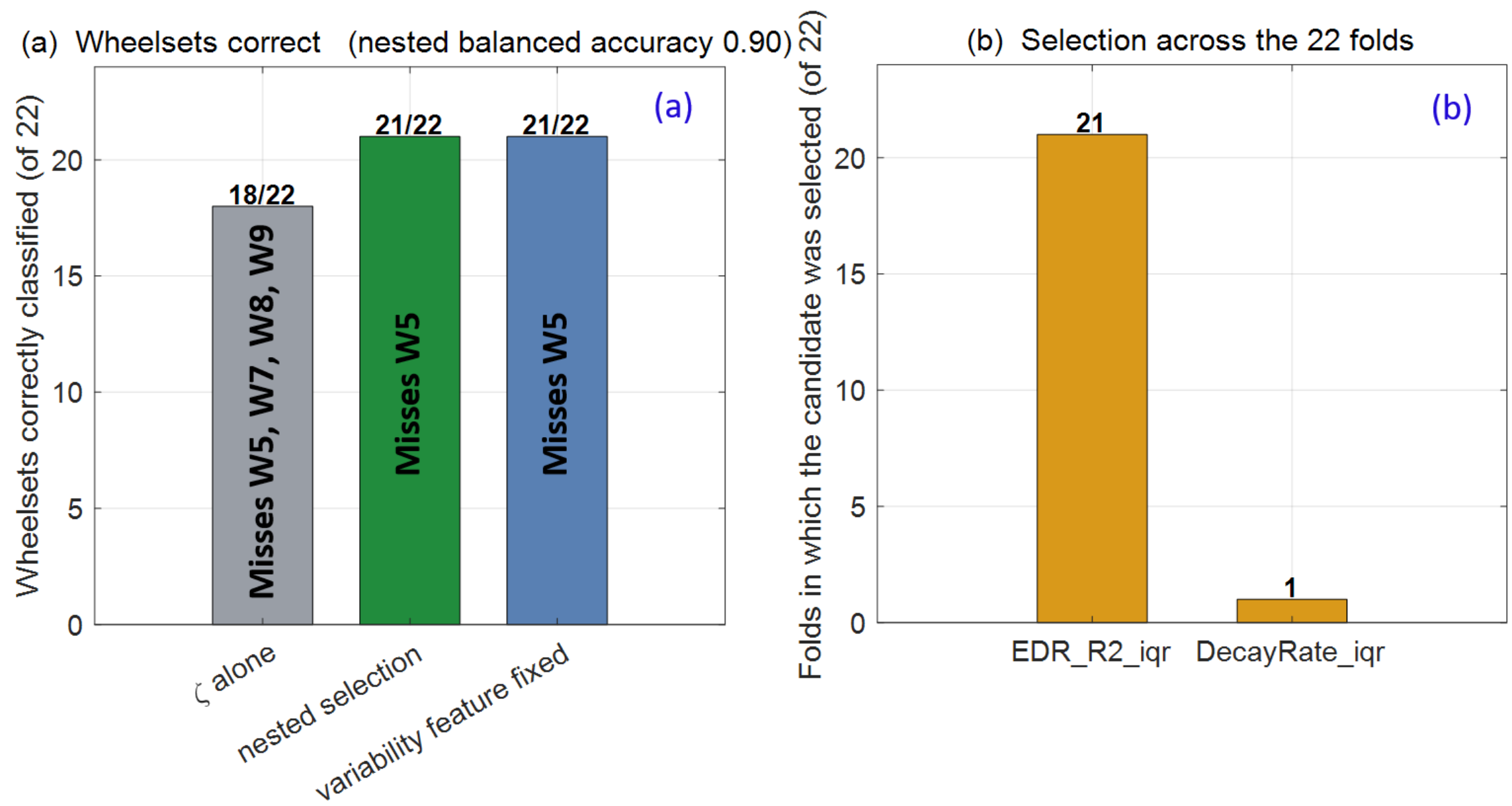

Figure 10. Nested hidden-wheel validation over the 22 wheelsets. (a) Wheelsets correctly classified by the damping ratio alone, by nested selection of the second feature, and with the second feature fixed in advance. (b) Number of folds in which each candidate was selected.

## 5.6 Feature Selection from a Single Dataset

To test whether a transferable second feature could be selected from a single dataset, feature selection and threshold calibration were restricted to that dataset before testing on the other. MxV selected Decay_R2_iqr, while Norfolk Southern selected DecayRate_iqr. Each resulting detector correctly classified eight of the eleven wheelsets in the other dataset. By comparison, fixing the variability feature identified during development using both datasets yielded eleven correct classifications in both transfer directions (Figure 11). In the Norfolk Southern dataset, damping alone already separated the training wheelsets, and six second-feature candidates tied for the highest inner score. In the MxV dataset, the selected feature improved the inner validation score but still failed to transfer. These results show that the available single-site data did not reliably identify a transferable second feature. The reasons for the different selections are discussed in Section 6.2.

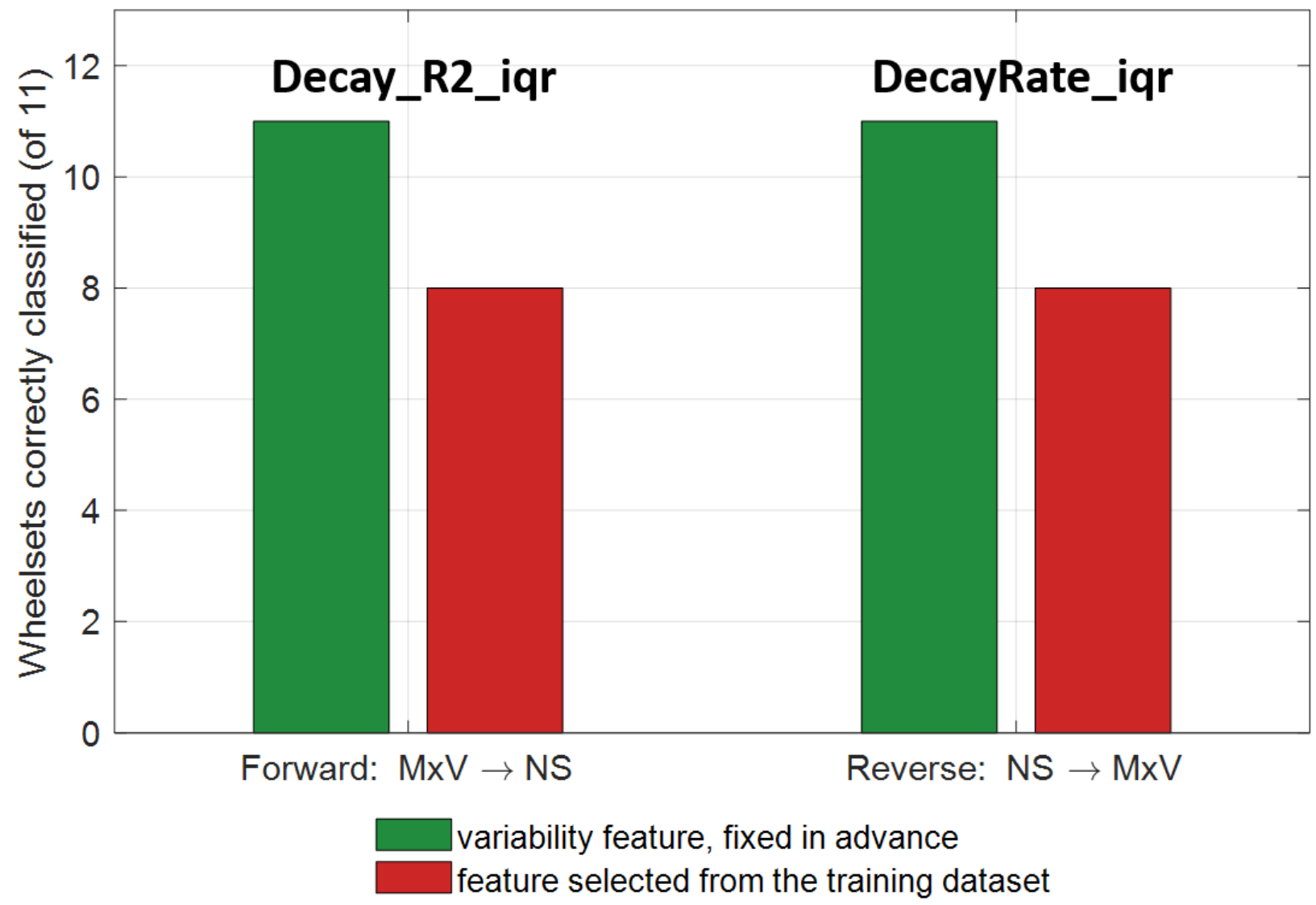

Figure 11. Cross-dataset classification with the variability feature fixed from pooled development and with the second feature selected using only the training dataset. Selection on MxV returned Decay_R2_iqr; selection on Norfolk Southern returned DecayRate_iqr.

Table 1. Evaluation results, with position-level and wheel-level scores reported separately.

| Evaluation | Level | Healthy recall | Defective recall | Balanced accuracy | Correct | Misclassified |
|---|---|---|---|---|---|---|
| 109 features, pooled leave-one-wheel-out | position | 0.73 | 0.79 | 0.76 | — | — |
| 109 features, MxV to NS | position | 0.06 | 0.90 | 0.48 | — | — |
| 109 features, MxV to NS | wheel | 0.00 | 1.00 | 0.50 | 8 of 11 | W20, W21, W22 |
| Two features, MxV to NS | wheel | 1.00 | 1.00 | 1.00 | 11 of 11 | none |
| Two features, NS to MxV | wheel | 1.00 | 1.00 | 1.00 | 11 of 11 | none |
| Damping ratio alone, leave-one-wheel-out | wheel | 0.80 | 0.82 | 0.81 | 18 of 22 | W5, W7, W8, W9 |
| Two features, feature fixed, leave-one-wheel-out | wheel | 0.80 | 1.00 | 0.90 | 21 of 22 | W5 |
| Two features, feature selected in each fold | wheel | 0.80 | 1.00 | 0.90 | 21 of 22 | W5 |
| Single-dataset selection, MxV to NS | wheel | 0.00 | 1.00 | 0.50 | 8 of 11 | W20, W21, W22 |
| Single-dataset selection, NS to MxV | wheel | 0.50 | 0.78 | 0.64 | 8 of 11 | W5, W7, W8 |
| Damping ratio alone, MxV to NS | wheel | 1.00 | 1.00 | 1.00 | 11 of 11 | None |
| Damping ratio alone, NS to MxV | wheel | 1.00 | 0.67 | 0.83 | 8 of 11 | W7, W8, W9 |

# 6. Discussion

## 6.1 Interpretation of the Two Features

The classification results show that the two features provide complementary information about the measured ring-down. Some defective wheelsets were identified by changes in effective damping, while others were identified by changes in decay-fit consistency across repeated strikes. The benefit of combining the features therefore stems from their responses to different decay patterns. The physical mechanisms responsible for these patterns were not isolated in this study. The features should therefore be treated as screening indicators rather than direct measurements of a specific defect mechanism. Differences in wheel geometry, support condition, impact excitation, and measurement conditions may also affect the recorded response.

Both features can be calculated from the current impact recordings without requiring an earlier measurement of the same wheel or its dimensions as inputs. This is useful when the service history and undamaged response of a wheel are unavailable. However, the absence of geometry as an input does not establish that the features are independent of wheel geometry. The detector provides a referral for further inspection; it does not determine defect type, size, or location. A defect could also remain undetected if both features remain within the ranges represented by the healthy training wheelsets.

### 6.2 Cross-Dataset Performance and Feature Selection

The pooled and cross-dataset evaluations address different questions. In pooled leave-one-wheel-out validation, the test wheelset was hidden, but wheelsets from both sites remained in the training data. In cross-dataset testing, the entire test site was excluded from training. The poor transfer of the 109-feature model shows that good performance within a combined population does not necessarily translate to performance at a new site. This behavior is consistent with population-based structural health monitoring studies showing that differences between measurement populations can affect feature transfer [13], [14].

The results from the two-feature detector must be interpreted carefully. Once the damping and variability features were chosen, thresholds calibrated on one dataset worked when applied to the other. However, the variability feature was originally identified using information from both datasets. The cross-dataset test therefore confirms that the thresholds transferred after the two features were fixed. It does not show that the same features would have been selected using data from only one site. This limitation was examined by repeating feature selection using one dataset at a time. Each dataset selected a different second feature, and neither selected feature transferred as well as the variability feature. In the Norfolk Southern dataset, damping already separated the training wheelsets, so there was little information available to identify a useful second feature. In the MxV dataset, the selected feature improved performance within the training data but did not perform as well on the other dataset.

Nested validation showed that the variability feature was consistently selected when wheelsets from both sites were available and each test wheelset was withheld. However, the overall detector was still developed using information from the two available datasets. A third independent dataset is therefore needed to confirm both the feature selection and the calibrated thresholds.

### 6.3 Threshold Stability and Screening Decisions

The damping threshold was relatively stable across the available training sets. The independently calibrated values differed by approximately one percent, and the fixed-feature leave-one-wheel-out values spanned a range of 2.8%. The variability threshold changed more, differing by approximately one-third between the two site calibrations and by 63% across the leave-one-wheel-out folds. Although these changes did not alter the cross-dataset classifications in this study, they show that the variability threshold is more sensitive to the wheelsets used for calibration. The two thresholds should therefore continue to be assessed separately as the dataset grows.

Signal quality and fit quality play different roles in the detector. The damping estimate requires a strong signal and an adequate exponential fit. The variability feature uses changes in fit quality as the measured quantity. Removing every strike with a poor exponential fit would therefore remove part of the variation that the second feature is intended to capture.

Weak recordings do not appear to explain the elevated variability in the three MxV wheelsets identified by the second feature. Their median signal-to-noise ratios ranged from 53.0 to 55.8 dB, compared with 51.7 to 54.0 dB for the healthy MxV wheelsets. All were above the signal-quality requirement, and the correlation between signal-to-noise ratio and decay-fit quality was +0.05. This evidence argues against low signal level as a simple explanation, but it does not identify whether the variation arose from structural condition, impact repeatability, interaction between vibration modes, or another source. The OR rule and the treatment of missing measurements intentionally favor referral for further inspection. Either feature can add a position flag, and a position with neither feature available is also flagged. These choices support defective-wheel detection but can increase the number of healthy wheelsets referred. Future evaluations should therefore report healthy-class recall, missing-measurement frequency, and the resulting inspection referral rate.

### 6.4 Limitations

The study included 22 wheelsets, of which only five were healthy, and all five healthy wheelsets were new. Therefore, the healthy class does not represent worn wheels that remain acceptable for service. With so few healthy examples, a single classification error changes healthy-class recall by 0.20 and balanced accuracy by 0.10. Additional healthy wheelsets, particularly worn but serviceable ones, are needed to estimate the referral rate and calibrate the thresholds more reliably.

Support conditions were also confounded with wheel conditions in the Norfolk Southern dataset. The healthy wheelsets were tested on rail, whereas the defective wheelsets were tested on the ground. Transfer from the MxV dataset does not resolve this uncertainty because a support-related response could reinforce the observed separation between the classes. To evaluate this effect, healthy and defective wheelsets must be measured under both support conditions.

Finally, the detector provides a binary screening decision. It does not identify defect type, location, or severity. The defect-specific recall values reported for the 109-feature model reflect binary screening performance within those groups and should not be interpreted as defect classification. Further controlled testing and prospective evaluation on an independent dataset are required before broader application.

## 7. Conclusions

This study evaluated an air-coupled ultrasonic method for screening railway wheelsets after hammer excitation. The detector combines effective damping with variation in decay-fit quality across repeated strikes. Both features are computed from the current recordings and do not require a prior measurement of the same wheel. The results show that the two features provide complementary information about the measured ring-down.

The 109-feature Random Forest model performed better when wheelsets from both sites were included in the training data, but its performance decreased when applied to a dataset from a different site. In comparison, once the two decay features were fixed, thresholds calibrated on either dataset correctly classified all eleven wheelsets in the other dataset. Nested validation, in which the second feature was selected without the test wheelset, correctly classified 21 of the 22 wheelsets.

Feature selection was less reliable when restricted to a single dataset. The resulting detectors correctly classified only eight of the eleven wheelsets in the other dataset. The present results therefore support transferring the calibrated thresholds after the two features have been fixed, but they do not yet confirm that the same features can be selected independently from a single-site dataset.

The study was limited to 22 stationary wheelsets from two sites. Only five wheelsets were healthy, and all were new wheels. Further evaluation should use an independent dataset that includes worn but serviceable wheels, as well as both healthy and defective wheels, under controlled support conditions. Applying the method to moving wheels will also require additional work on excitation, background noise, and defining repeated measurements.